\documentclass[aps,prd,preprint,superscriptaddress]{revtex4-1}
\usepackage[mathscr]{euscript}
\usepackage{multirow}
\usepackage{array}
\usepackage{booktabs}
\usepackage{threeparttable}
\usepackage{float,epsfig}
\usepackage{color,soul}
\usepackage[normalem]{ulem}
\usepackage{bm}% bold math
\usepackage{graphicx}% Include figure files
\usepackage{subfigure}
\usepackage{appendix}
\usepackage{amsmath,amssymb,amsthm}
\usepackage[colorlinks=true,linkcolor=blue,urlcolor=blue]{hyperref}
\newcommand{\bea}{\begin{eqnarray}}
\newcommand{\eea}{\end{eqnarray}}
\newcommand{\beq}{\begin{equation}}
\newcommand{\eeq}{\end{equation}}

\def\/{\over}
\allowdisplaybreaks[4]

\begin{document}

%\preprint{}

\title{Repulsive quantum gravitoelectric-gravitomagnetic interaction}
%\title{Quantum gravitational interaction induced by quantized gravitomagnetic fields}

\author{Di Hao}
\affiliation{Department of Physics, Synergetic Innovation Center for Quantum Effects and Applications,\\ and Institute of Interdisciplinary Studies, \\Hunan Normal University, Changsha, Hunan 410081, China}
\author{Jiawei Hu}
\email[Corresponding author. ]{jwhu@hunnu.edu.cn}
\affiliation{Department of Physics, Synergetic Innovation Center for Quantum Effects and Applications,\\ and Institute of Interdisciplinary Studies, \\Hunan Normal University, Changsha, Hunan 410081, China}
\author{Hongwei Yu}
\email[Corresponding author. ]{hwyu@hunnu.edu.cn}
\affiliation{Department of Physics, Synergetic Innovation Center for Quantum Effects and Applications,\\ and Institute of Interdisciplinary Studies, \\Hunan Normal University, Changsha, Hunan 410081, China}

\begin{abstract}

We investigate, in the framework of linearized quantum gravity, the quantum gravitational interaction between a gravitoelectrically polarizable object and a gravitomagnetically polarizable object. This interaction originates from the coupling between the instantaneous mass quadrupole moment and the mass-current quadrupole moment of the objects, induced by fluctuating gravitoelectric and gravitomagnetic fields in a vacuum. Using leading-order perturbation theory, we derive the explicit expression of the quantum gravitoelectric-gravitomagnetic interaction energy, which shows a distance dependence of $r^{-8}$ in the near regime and $r^{-11}$ in the far regime, where $r$ is the distance between the two objects. Remarkably, this interaction energy is positive, indicating that the force is repulsive. Since interactions between objects polarizable in the same gravitoelectric or gravitomagnetic manner are inherently attractive, for  objects which are both gravitoelectrically and gravitomagnetically polarizable, the overall quantum gravitational interaction potential is reduced  when the repulsive quantum gravitoelectric-gravitomagnetic interaction is taken into account. However, for two isotropically polarizable objects with identical gravitoelectric and gravitomagnetic polarizabilities and energy level spacing, the repulsive quantum interaction cannot surpass the attractive interactions.

\end{abstract}

%\pacs{}
%\keywords{}

\maketitle

\section{Introduction}
\label{sec_in}
\setcounter{equation}{0}
%%%%%%%%%%%%%%%%%%%%%%%%%%%%%%%%%%%%%%%%%%%%%%%%%%
Gravitational waves, ripples in spacetime predicted by general relativity~\cite{Einstein}, were not directly confirmed until the Laser Interferometer Gravitational-Wave Observatory (LIGO) detected signals from black hole mergers~\cite{LIGO}. This detection measured their tiny effects on the length differences between the arms of the interferometer. More recently, this prediction was further validated by the detection of nanohertz stochastic background gravitational wave signals, observed through their influence on the arrival times of radio pulses from arrays of pulsars~\cite{CPTA,PPTA,EPTA,NANO}. These discoveries highlight the classical effects of gravity described by general relativity.

Naturally, one may wonder what happens if, like electromagnetic waves, gravitational waves can also be quantized. Unfortunately, a complete theory of quantum gravity remains elusive. Nevertheless, quantum gravitational phenomena at low-energy scales can still be explored. For example, by treating general relativity as an effective field theory, quantum corrections to the classical Newtonian gravitational interaction between a pair of mass monopoles have been investigated by summing one-loop Feynman diagrams involving off-shell gravitons~\cite{Donoghue1994prl,Donoghue1994prd,Hamber1995,Kirilin2002,Holstein2003,Holstein2005}.

If gravity is indeed quantum, a natural consequence would be the inevitable existence of fluctuating background gravitational fields, even in a vacuum. These gravitational quantum vacuum fluctuations, representing typical low-energy quantum gravitational effects, can be studied using linearized quantum gravity, where the gravitational field is treated as a linear perturbation on flat spacetime. In weak gravity scenarios like the one considered here, the gravitational field equations can be expressed in a form analogous to Maxwell's equations, known as Weyl gravitoelectromagnetism~\cite{Campbell1976,Matte1953,Campbell1971,Szekeres1971,Maartens1998,Ruggiero2002,Ramos2010}. In this formalism, the gravitational field is decomposed into two parts: the gravitoelectric field and the gravitomagnetic field, analogous to the electric and magnetic fields, respectively.

Any nonpointlike object would then be polarized by these fluctuating gravitational fields, inducing instantaneous mass quadrupole moments from the gravitoelectric field and mass-current quadrupole moments from the gravitomagnetic field. Consequently, quantum corrections to the classical Newtonian gravitational interaction arise from interactions between these instantaneous quadrupole moments. In this context, quantum corrections due to the gravitational interaction between instantaneous mass quadrupole moments induced by fluctuating gravitoelectric fields have been studied both between a nonpointlike object and a boundary~\cite{Hu2017}, and between two nonpointlike objects~\cite{Ford2016,Wu2016,Holstein2017}. Subsequently, these effects have been explored in several scenarios, including the quantum gravitational interaction between two nonpointlike objects induced by a thermal bath of gravitons~\cite{Wu2017}, and the interaction between two nonpointlike objects near a gravitational boundary~\cite{yu2018}. Additionally, studies have investigated the interaction between a pair of nonpointlike objects in symmetric and antisymmetric entangled states~\cite{yongs2020epjc}, as well as the nonadditive interaction among three nonpointlike objects~\cite{yongs2022prd}.

Recently, quantum corrections due to the gravitational interaction between instantaneous mass-current quadrupole moments induced by fluctuating gravitomagnetic fields in a vacuum have also been investigated~\cite{hao2024}. Interestingly, the quantum gravitational interactions between two gravitoelectrically polarizable objects~\cite{Ford2016,Wu2016,Holstein2017} and between two gravitomagnetically polarizable objects~\cite{hao2024} 
both exhibit an  $r^{-10}$ dependence in the nonretarded regime and an $r^{-11}$ dependence in the retarded regime, and are both inherently attractive.

Therefore, a gap remains in our understanding of the quantum gravitational interaction between two nonpointlike objects arising from the instantaneous quadrupole moments induced by gravitational vacuum fluctuations, specifically, the quantum corrections due to the gravitational interaction between an instantaneous mass quadrupole moment and a mass-current quadrupole moment. Moreover, it is particularly interesting to determine whether this interaction leads to an attractive force that enhances the classical gravitational interaction, or a repulsive force that causes a suppression, given that previous quantum corrections involving mass and mass-current quadrupole moments have resulted in attractive forces.

To complete this final piece of the puzzle, we now investigate the quantum gravitational interaction energy between a gravitoelectrically polarizable object and a gravitomagnetically polarizable object. Specifically, we focus on the distance dependence of this interaction energy and whether the interaction is repulsive or attractive.

The paper is organized as follows. First, we derive the quantum gravitational energy resulting  from the crossed gravitoelectric-gravitomagnetic term using leading-order perturbation theory. Then, we examine its asymptotic behavior in both the near and far regimes, and analyze the sign of the energy to determine whether the interaction is repulsive or attractive. Throughout the paper, we adopt Greek letters for four-dimensional spacetime indices (0-3), and Latin letters for three-dimensional spatial indices (1-3). We also employ the Einstein summation convention for repeated indices. Unless otherwise specified, natural units $c = \hbar = 1$ are adopted.

%%%%%%%%%%%%%%%%%%%%%%%%%%%%%%%%%%%%%%%%%%%%%%%%%%
\section{The basic formalism}
\label{sec_ge}
%\setcounter{equation}{0}
%%%%%%%%%%%%%%%%%%%%%%%%%%%%%%%%%%%%%%%%%%%%%%%%%%
Under the weak-field approximation, the spacetime metric $g_{\mu\nu}$ can be expressed as a linearized perturbation $h_{\mu\nu}$ propagating on a flat background spacetime $\eta_{\mu\nu}$, i.e., $g_{\mu\nu}=\eta_{\mu\nu}+h_{\mu\nu}$. Within the framework of linearized quantum gravity, $h_{\mu\nu}$ is quantized, and in the transverse-traceless (TT) gauge, it takes the standard form~\cite{Wu2017,Oniga2016}
\beq \label{hij_new}
h_{ij}(\textbf{r},t)
=\sum_{\textbf{k},\lambda} \sqrt{\frac{8\pi G}{(2\pi)^3\omega}} e_{ij}(\textbf{k},\lambda) 
\left[ g_{\textbf{k},\lambda}
e^{i\textbf{k}\cdot\textbf{r}-i\omega t}
+g_{\textbf{k},\lambda}^{\dagger} e^{-i\textbf{k}\cdot\textbf{r}+i\omega t}\right],
\eeq
where $G$ is the Newton's gravitational constant, $\textbf{k}$  the wave vector, $\omega=|\textbf{k}|$ the frequency, $e_{ij}(\textbf{k},\lambda)$ the gravitational polarization tensor, $\lambda$ the polarization states, and $g_{\textbf{k},\lambda}$ and $g_{\textbf{k},\lambda}^{\dagger}$ the annihilation and creation operators of the gravitational field with wave vector $\textbf{k}$ and polarization $\lambda$ satisfying the following commutation relations:
\bea \label{comm}
\nonumber\left[ g_{\textbf{k},\lambda},g_{\textbf{k}\prime,\lambda \prime} \right] =0,  \\
\nonumber\left[ g_{\textbf{k},\lambda}^{\dagger},g_{\textbf{k}\prime,\lambda \prime}^{\dagger} \right] =0, \\
\left[ g_{\textbf{k},\lambda},g_{\textbf{k}\prime,\lambda \prime}^{\dagger} \right] =\delta _{\lambda \lambda \prime}\delta _{\textbf{kk}\prime},
\eea
with $\delta_{ab}$ being the Kronecker symbol.

The fluctuating gravitational fields can induce quadrupole moments in a nonpointlike object, and the interaction Hamiltonian between the fluctuating gravitational fields and the induced quadrupole moments can be written as~\cite{hao2024}
\bea
    \label{object_field_Hamiltonian}
    H_{\mathrm{int}}&=-\frac{1}{2}Q^{ij}E_{ij}-\frac{1}{3}S^{ij}B_{ij},
\eea
where $Q^{ij}$ represents the induced mass quadrupole moment, and $S^{ij}$ denotes the induced mass-current quadrupole moment of the object. Here, $E_{ij}$ and $B_{ij}$ are the gravitoelectric and gravitomagnetic fields, respectively, which are defined as~\cite{Campbell1976,Matte1953,Campbell1971,Szekeres1971,Maartens1998,Ruggiero2002,Ramos2010}
\beq \label{my_Eij}
E_{ij}=-C_{0i0j},
\eeq
and  
\beq \label{Bij}
B_{ij}=\frac{1}{2}\epsilon_{ifl}C_{fl0j},
\eeq
through an analogy between the Maxwell equations and the linearized gravitational field equations under the weak-field approximation, where $C_{\alpha\beta\gamma\sigma}$ is the Weyl tensor, and $\epsilon_{ifl}$ is the third-order Levi-Civita tensor. The object-field interaction Hamiltonian given by Eq.~(\ref{object_field_Hamiltonian}) describes the gravitational quadrupolar coupling. Specifically, the explicit expressions for the two kinds of gravitational quadrupole moments take the standard form~\cite{hao2024,Flanagan2007,Rezzolla1999}
\beq \label{Q_ex}
Q_{jk}=\int d^{3}x\rho(x)(x_{j}x_{k}-\frac{1}{3}\delta_{jk}r^{2}),
\eeq
and  
\beq \label{S_ex}
S_{lj}=\int d^{3}x\epsilon_{if(l}x_{j)}x_{i}\rho (x)v_{f},
\eeq
where $\rho(x)$ is the mass density of the nonpointlike object, and $\rho(x) v_{f}$ denotes the localized mass-current density. Note that $A_{(ij)}=\frac{1}{2}(A_{ij}+A_{ji})$.

Furthermore, based on Weyl gravitoelectromagnetism and the theory of linearized quantum gravity, the explicit expressions for the quantized gravitoelectric and gravitomagnetic fields can be derived as follows. Under the weak-field approximation, the Riemann curvature tensor $R_{\alpha\beta\mu\nu}$ can be written in terms of the gravitational perturbation $h_{\mu\nu}$ as
\beq \label{R_tensor}
R_{\alpha\beta\mu\nu}=\frac{1}{2}\left(\partial_{\beta}\partial_{\mu}h_{\alpha\nu}-\partial_{\alpha}\partial_{\mu}h_{\beta\nu}-\partial_{\beta}\partial_{\nu}h_{\alpha\mu}+\partial_{\alpha}\partial_{\nu}h_{\beta\mu}\right).
\eeq
Then, according to the definitions of $E_{ij}$ and $B_{ij}$ shown in Eqs. (\ref{my_Eij}) and (\ref{Bij}), and noting that the Riemann curvature tensor is equivalent to the Weyl tensor in the vacuum case, the gravitoelectric and gravitomagnetic fields can be expressed as
\bea \label{Eij_h}
E_{ij}
=-\frac{1}{2}\ddot{h}_{ij},
\eea 
and
\bea \label{Bij_h}
B_{ij}
=-\frac{1}{2}\epsilon_{ifl}\partial_{f}\dot{h}_{lj},
\eea
respectively, where a dot denotes the first derivative with respect to time. Substituting Eq.~(\ref{hij_new}) into Eqs. (\ref{Eij_h}) and (\ref{Bij_h}), the quantized fields take the form
\bea \label{q_Eij_h}
E_{ij}(\textbf{r},t)
=-\frac{1}{2}\sum_{\lambda}\int d^{3}\textbf{k}\sqrt{\frac{8\pi G\omega^{3}}{(2\pi)^{3}}}e_{ij}(\textbf{k},\lambda)\left[
g_{\textbf{k},\lambda}(t)
e^{i\textbf{k}\cdot\textbf{r}}
+g_{\textbf{k},\lambda}^{\dagger}(t)
e^{-i\textbf{k}\cdot\textbf{r}}\right],
\eea
and
\bea \label{q_Bij_h}
B_{ij}(\textbf{r},t)
=-\frac{1}{2}\sum_{\lambda}\int d^{3}\textbf{k}\sqrt{\frac{8\pi G\omega^{3}}{(2\pi)^{3}}}\epsilon_{ifl}e^{f}_{3}e_{lj}(\textbf{k},\lambda)\left[
g_{\textbf{k},\lambda}(t)
e^{i\textbf{k}\cdot\textbf{r}}
+g_{\textbf{k},\lambda}^{\dagger}(t)
e^{-i\textbf{k}\cdot\textbf{r}}\right],
\eea
respectively. Here, $g_{\textbf{k},\lambda}(t)=g_{\textbf{k},\lambda}e^{-i\omega t}$ and $g^\dagger_{\textbf{k},\lambda}(t)=g^\dagger_{\textbf{k},\lambda}e^{i\omega t}$.  $\textbf{e}_{3}=\textbf{k}/|\textbf{k}|$ represents the unit vector along the propagation direction of the gravitational field, and $e^{f}_{3}~(f=x,y,z)$ denotes the $f$th coordinate component of $\textbf{e}_{3}$.

\section{The quantum gravitoelectric-gravitomagnetic interaction}\label{sec3}

We consider a pair of gravitationally polarizable objects A and B coupled to fluctuating gravitational fields in a vacuum. Both objects can be modeled as two-level systems, with the excited and ground states being $|e_{A(B)}\rangle$ and $|g_{A(B)}\rangle$ respectively, and the corresponding energy level spacing is $\omega_{A(B)}$. The total Hamiltonian of the object-field system can thus be written as
\beq \label{tot_Hamiltonian}
H_{\mathrm{tot}}=H_{A}+H_{B}+H_{F}+H_{AF}+H_{BF},
\eeq
where $H_{F}$ is the Hamiltonian of the fluctuating gravitational fields, $H_{A(B)}$ is the Hamiltonian of object A(B), and $H _{A(B)F}$ represents the interaction between object A(B) and fluctuating gravitational fields. 
For simplicity, we assume that object A is gravitoelectrically polarizable while object B is gravitomagnetically polarizable; hence, the interaction Hamiltonian of the system reads
\beq \label{H_GE_GM}
H^{\mathrm{cross}}_{\mathrm{int}}=H_{AF}+H_{BF}=-\frac{1}{2}Q^{ij}_{A}E_{ij}(\textbf{r}_{A})-\frac{1}{3}S^{ij}_{B}B_{ij}(\textbf{r}_{B}).
\eeq

We assume that both objects are initially in their ground states, $|g_{A}\rangle$ and $|g_{B}\rangle$, and  the gravitational field is in the vacuum state $|0\rangle$. Therefore, the initial state of the whole system can be expressed as
\beq \label{state_initial}
|\phi\rangle=|g_{A}\rangle|g_{B}\rangle|0\rangle.
\eeq
Then, the quantum gravitational interaction between the two objects induced by gravitational vacuum fluctuations 
can be obtained using fourth-order perturbation theory, which takes the standard form
\beq \label{E_AB}
\Delta E^{\mathrm{cross}}_{AB}=-\sum_{\mathrm{I},\mathrm{II},\mathrm{III}\neq\phi}\frac{\langle \phi|H^{\mathrm{cross}}_{\mathrm{int}}|\mathrm{I}\rangle\langle \mathrm{I}|H^{\mathrm{cross}}_{\mathrm{int}}|\mathrm{II}\rangle\langle \mathrm{II}|H^{\mathrm{cross}}_{\mathrm{int}}|\mathrm{III}\rangle\langle \mathrm{III}|H^{\mathrm{cross}}_{\mathrm{int}}|\phi\rangle}{(E_{\mathrm{I}}-E_{\phi})(E_{\mathrm{II}}-E_{\phi})(E_{\mathrm{III}}-E_{\phi})}.
\eeq
Here, $| \mathrm{I}\rangle$, $| \mathrm{II}\rangle$ and $| \mathrm{III} \rangle$ are the three intermediate states in the interaction processes. See Table \ref{24I} in Appendix~\ref{appd1} for all possible intermediate states and the corresponding energy denominators in Eq. (\ref{E_AB}).

Taking case (1) in Table \ref{24I} as an example. Substituting $|\mathrm{I}\rangle$ , $|\mathrm{II}\rangle$ and $|\mathrm{III}\rangle$ into Eq. (\ref{E_AB}) yields
\bea \label{E_cross(1)}
\nonumber
\Delta E^{\mathrm{cross}}_{AB(1)}(\textbf{r}_{A},\textbf{r}_{B})&=&-\frac{1}{36}\int_{0}^{+\infty}d\omega \int_{0}^{+\infty}d{\omega}'\frac{1}{\mathcal{D}_{\mathrm{(1)}}}\left[\hat{Q}^{ij}_{A}\hat{S}^{ab}_{B}G^{\mathrm{E-M}}_{ijab}\left(\omega,\textbf{r}_{A},\textbf{r}_{B}\right)\right]  \\
&&\times\left[\hat{Q}^{*kl}_{A}\hat{S}^{*cd}_{B}G^{\mathrm{E-M}}_{klcd}\left({\omega}',\textbf{r}_{A},\textbf{r}_{B}\right) \right].  
\eea
Here $\mathcal{D}_{(1)}$ is the energy denominator corresponding to case (1), $\hat{Q}^{ij}_{A}=\langle g_{A}|Q^{ij}_{A}|e_{A}\rangle$ is the mass quadrupole transition matrix element, and $\hat{S}^{ij}_{B}=\langle g_{B}|S^{ij}_{B}|e_{B}\rangle$ is the mass-current quadrupole transition matrix element. The conjugate terms are denoted as $\hat{Q}^{*ij}_{A}$ and $\hat{S}^{*ij}_{B}$, respectively. $G^{\mathrm{E-M}}_{ijab}(\omega, \textbf{r}_{A},\textbf{r}_{B})$ denotes the crossed two-point correlation function of the gravitoelectric and gravitomagnetic fields in the frequency domain, which can be expressed as
\beq \label{G_ijab}
G^{\mathrm{E-M}}_{ijab}(\omega, \textbf{r}_{A},\textbf{r}_{B})=\langle 0|E_{ij}(\omega,\textbf{r}_{A})B_{ab}(\omega,\textbf{r}_{B})|0\rangle.
\eeq

Next, we organize the contributions from the remaining 11 intermediate processes into the form of Eq. (\ref{E_cross(1)}), respectively, and then sum up all the contributions from the 12 intermediate processes. Generally, for a nonpointlike object that obeys time-reversal symmetry, the eigenstates of the object's Hamiltonian in the position representation can be chosen as real functions \cite{Buhmann2013}. Hence, we consider objects with real space-part wave functions $\psi _n\left( \mathbf{r} \right)$, where $n$ denotes the quantum number of the energy eigenstate. 
On the other hand, the nonpointlike object considered here can be modeled as a system composed of several mass points $m_{\mathbb{P}}$ $(\mathbb{P}\in \mathrm{A\ or\ B} )$. Then, the mass density distribution $\rho(x)$ in the quadrupole moments Eqs.~(\ref{Q_ex}) and (\ref{S_ex}) takes the form of $\sum_{\mathbb{P} }{m_{\mathbb{P}}\delta^{(3)}(x-x_{\mathbb{P}})}$, where $\delta^{(3)}(x)$ is the three-dimensional Dirac delta function. Therefore, the mass and mass-current quadrupole transition matrix elements can be expressed as
\bea \label{re1}
\langle n|Q_{A}^{ij}|m\rangle 
= \int{d^3r_k}\psi _{n}^{*}\left( r_k \right) \left[ \sum_{\mathbb{P} \in \mathrm{A}}{m_{\mathbb{P}}\left( x_{i}^{\mathbb{P}}x_{j}^{\mathbb{P}}-\frac{1}{3}\delta _{ij}r_{\mathbb{P}}^{2} \right)} \right] \psi _m\left( r_k \right), 
\eea
and
\bea \label{im2}
\langle n|S_{B}^{ij}|m\rangle 
=\int{d^3r_k}\psi _{n}^{*}\left( r_k \right) \left[ \sum_{\mathbb{P} \in \mathrm{B}}{\epsilon _{kl\left( i \right.}x^{\mathbb{P}}_{\left. j \right)}x^{\mathbb{P}}_{k} ( -i\hbar\nabla^{\mathbb{P}}_{l}}) \right] \psi _m\left( r_k \right),
\eea
respectively. Note that the momentum $m_{\mathbb{P}} v^{\mathbb{P}}_{l}$ of mass point $\mathbb{P}$ in object B has been replaced by the quantum operator $-i\hbar\nabla^{\mathbb{P}}_{l}$. From Eqs. (\ref{re1}) and (\ref{im2}), it is clear that the mass quadrupole transition matrix elements are purely real, satisfying $\hat{Q}^{ij}_{A}=\hat{Q}^{*ij}_{A}$, while the mass-current quadrupole transition matrix elements are purely imaginary, leading to $\hat{S}^{ij}_{B}=-\hat{S}^{*ij}_{B}$. 
Based on this, and  with the help of the symmetry of the two-point function (see Appendix~\ref{appd2} for the proof)
\bea
    \label{Green_symmetry}
    G^{\mathrm{E-M}}_{ijkl}(\textbf{r},{\textbf{r}}',t,t')=-\langle 0|B_{ij}(\textbf{r},t)E_{kl}(\textbf{r}',t')|0\rangle=-G^{\mathrm{M-E}}_{ijkl}(\textbf{r},{\textbf{r}}',t,t'),
\eea
we  find, after an inspection of all the 12 possible intermediate processes, that the contributions to the interaction energy (\ref{E_AB}) corresponding to cases (2), (3), (7), (8), and (9) can be organized in the same form as Eq. (\ref{E_cross(1)}) and carry the same sign. In contrast, the contributions from cases (4), (5), (6), (10), (11), and (12) can also be organized in the same form as Eq. (\ref{E_cross(1)}) but carry the opposite sign. Taking this into account, the total interaction energy between the objects is obtained as follows:
\bea \label{E_cross_tot}
\nonumber
\Delta E^{\mathrm{cross}}_{AB}(\textbf{r}_{A},\textbf{r}_{B})&=&-\frac{1}{36}\int_{0}^{+\infty}d\omega \int_{0}^{+\infty}d{\omega}'\Big(\frac{1}{\mathcal{D}_{(1)}}+\frac{1}{\mathcal{D}_{(2)}}+\frac{1}{\mathcal{D}_{(3)}}+\frac{1}{\mathcal{D}_{(7)}}+\frac{1}{\mathcal{D}_{(8)}}+\frac{1}{\mathcal{D}_{(9)}}  \\
\nonumber&&-\frac{1}{\mathcal{D}_{(4)}}-\frac{1}{\mathcal{D}_{(5)}}-\frac{1}{\mathcal{D}_{(6)}}-\frac{1}{\mathcal{D}_{(10)}}-\frac{1}{\mathcal{D}_{(11)}}-\frac{1}{\mathcal{D}_{(12)}}\Big)  \\
\nonumber&&\times\left[\hat{Q}^{ij}_{A}\hat{S}^{ab}_{B}G^{\mathrm{E-M}}_{ijab}\left(\omega,\textbf{r}_{A},\textbf{r}_{B}\right)\right]\left[\hat{Q}^{*kl}_{A}\hat{S}^{*cd}_{B}G^{\mathrm{E-M}}_{klcd}\left({\omega}',\textbf{r}_{A},\textbf{r}_{B}\right) \right]  \\
\nonumber&=&-\frac{1}{36}\int_{0}^{+\infty}d\omega \int_{0}^{+\infty}d{\omega}'\hat{Q}^{ij}_{A}\hat{Q}^{*kl}_{A}\hat{S}^{ab}_{B}\hat{S}^{*cd}_{B} G^{\mathrm{E-M}}_{ijab}(\omega,\textbf{r}_{A},\textbf{r}_{B})G^{\mathrm{E-M}}_{klcd}({\omega}',\textbf{r}_{A},\textbf{r}_{B})  \\
&&\times\frac{4\left(\omega_{A}+\omega_{B}+\omega\right)}{\left(\omega_{A}+\omega_{B}\right)\left(\omega_{A}+\omega)(\omega_{B}+\omega\right)}\left(\frac{1}{\omega+{\omega}'}-\frac{1}{{\omega}'-\omega}\right).
\eea

For simplicity, we assume that both objects are isotropically polarizable. As a result, the mass and mass-current quadrupole transition matrix elements satisfy
\beq \label{polar_1}
\hat{Q}^{ij}_{A}\hat{Q}^{*kl}_{A}=(\delta_{ik}\delta_{jl}+\delta_{il}\delta_{jk})\hat{\alpha}_{A},
\eeq
and
\beq \label{polar_2}
\hat{S}^{ab}_{B}\hat{S}^{*cd}_{B}=(\delta_{ac}\delta_{bd}+\delta_{ad}\delta_{bc})\hat{\chi}_{B},
\eeq
respectively, where $\hat{\alpha}_A=|\hat{Q}^{ij}_A|^{2}$ and $\hat{\chi}_B=|\hat{S}^{ab}_B|^{2}$. Substituting Eqs.~(\ref{polar_1}) and (\ref{polar_2}) into Eq.~(\ref{E_cross_tot}), one can obtain
\bea \label{re_E_tot}
\nonumber
\Delta E^{\mathrm{cross}}_{AB}(\textbf{r}_{A},\textbf{r}_{B})&=&-\frac{1}{9(\omega_{A}+\omega_{B})}\int_{0}^{+\infty}d\omega \int_{0}^{+\infty}d{\omega}' G^{\mathrm{E-M}}_{ijab}(\omega,\textbf{r}_{A},\textbf{r}_{B})G^{\mathrm{E-M}}_{ijab}({\omega}',\textbf{r}_{A},\textbf{r}_{B})  \\
&&\times\frac{\hat{\alpha}_{A}\hat{\chi}_{B}\left(\omega_{A}+\omega_{B}+\omega\right)}{\left(\omega_{A}+\omega)(\omega_{B}+\omega\right)}\left(\frac{1}{\omega+{\omega}'}-\frac{1}{{\omega}'-\omega}\right).
\eea
The frequency-domain two-point function $G^{\mathrm{E-M}}_{ijab}(\omega,\textbf{r}_{A},\textbf{r}_{B})$ in the equation above can be derived from $G^{\mathrm{E-M}}_{ijab}(\textbf{r}_{A},\textbf{r}_{B},t_{A},t_{B})$ by a Fourier transform. The crossed two-point function of the gravitoelectric and gravitomagnetic fields can be obtained by using Eqs. (\ref{q_Eij_h}) and (\ref{q_Bij_h}) as
\bea \label{g_ijab_cal}
\nonumber
G^{\mathrm{E-M}}_{ijab}(\textbf{r},{\textbf{r}}',t,{t}')&=&\langle 0|E_{ij}(\textbf{r},t)B_{ab}(\textbf{r}',{t}')|0\rangle  \\
&=&\int d^{3}\textbf{k}\frac{G\omega^{3}}{(2\pi)^{2}}\mathcal{G}^{\mathrm{E-M}}_{ijab}(\textbf{k})e^{i\textbf{k}\cdot(\textbf{r}-{\textbf{r}}')-i\omega(t-{t}')},  
\eea
where $\mathcal{G}^{\mathrm{E-M}}_{ijab}(\textbf{k})$ represents the gravitoelectric-gravitomagnetic polarization summation term, which takes the form
\bea \label{polar_sum}
\nonumber
\mathcal{G}^{\mathrm{E-M}}_{ijab}(\textbf{k})&=&\epsilon_{apq}e^{p}_{3}\sum_{\lambda}e_{ij}(\textbf{k},\lambda)e_{qb}(\textbf{k},\lambda)   \\
&=&\epsilon_{iam}\hat{k}_m\delta_{jb}-\epsilon_{iam}\hat{k}_m\hat{k}_j\hat{k}_b+\delta_{ia}\epsilon _{jbn}\hat{k}_n-\hat{k}_i\hat{k}_a\epsilon _{jbn}\hat{k}_n,
\eea
with $\hat{k}_{i}$ being the $i$th coordinate component of the unit wave vector  $\hat{k}=\textbf{k}/k$. The detailed derivation of Eq. (\ref{polar_sum}) is shown in  Appendix~\ref{appd2}.
Let $\hat{k}_{x}=\sin\theta\cos\varphi$, $\hat{k}_{y}=\sin\theta\sin\varphi$, and $\hat{k}_{z}=\cos\theta$, i.e., transform the equation above to the spherical coordinate:
\beq \label{g_ijab}
\mathcal{G}^{\mathrm{E-M}}_{ijab}(\textbf{k})\xrightarrow{(\theta,\varphi)}\mathcal{G}^{\mathrm{E-M}}_{ijab}(\theta,\varphi),
\eeq
the time-domain two-point function becomes 
\beq \label{G_ijab_sph}
G^{\mathrm{E-M}}_{ijab}(r,\Delta t)=\int_{0}^{+\infty}d\omega\frac{G\omega^{5}}{(2\pi)^{2}}\int_{0}^{\pi}d\theta\sin\theta\int_{0}^{2\pi}d\varphi \mathcal{G}^{\mathrm{E-M}}_{ijab}(\theta,\varphi)e^{i\omega(r\cos\theta-\Delta t)},
\eeq
where $r=|\textbf{r}-{\textbf{r}}'|$ is the distance between the two objects, and $\Delta t=t-{t}'$. Performing the Fourier transform, the two-point function in the frequency domain is obtained as
\bea \label{G_ijab_fre}
\nonumber
G^{\mathrm{E-M}}_{ijab}(\bar{\omega},\textbf{r}_{A},\textbf{r}_{B})&=&\frac{1}{2\pi}\int_{-\infty}^{+\infty}d(\Delta t)e^{i\bar{\omega}\Delta t}G^{\mathrm{E-M}}_{ijab}(r,\Delta t)  \\
&=&\frac{G\bar{\omega}^{5}}{(2\pi)^{2}}\int_{0}^{\pi}d\theta\sin\theta\int_{0}^{2\pi}d\varphi \mathcal{G}^{\mathrm{E-M}}_{ijab}(\theta,\varphi)e^{i\bar{\omega} r\cos\theta}.
\eea
Then, by substituting Eq. (\ref{G_ijab_fre}) into Eq. (\ref{re_E_tot}) and performing the integration over the variables $(\theta,\varphi,{\theta}’,{\varphi}’)$, one obtains
\bea \label{E_AB_r}
\nonumber
\Delta E^{\mathrm{cross}}_{AB}(r)&=&-\frac{8G^{2}}{9\pi^{2}(\omega_{A}+\omega_{B})r^{8}}\int_{0}^{+\infty}\omega d\omega\int_{0}^{+\infty}{\omega}'d{\omega}'\frac{\hat{\alpha}_{A}\hat{\chi}_{B}(\omega_{A}+\omega_{B}+\omega)}{(\omega_{A}+\omega)(\omega_{B}+\omega)}       \\
&&\times\left(\frac{1}{\omega+{\omega}'}-\frac{1}{{\omega}'-\omega}\right)\Big[A_{1}(\omega r,{\omega}'r)\cos({\omega}'r)+B_{1}(\omega r,{\omega}'r)\sin({\omega}'r)\Big],
\eea
where
\bea \label{A_1}
\nonumber
A_{1}(x,x’)&=&xx'\left( -45-x'^2x^2+3x^2+3x'^2 \right) \cos x   \\
&&-x'\left( -45+18x^2-2x'^2x^2+3x'^2 \right) \sin x,
\eea
and
\bea \label{B_1}
\nonumber
B_{1}(x,x')&=&x\left( 45-18x'^2-3x^2+2x'^2x^2 \right) \cos x  \\
&&-\left( 45-18x^2-18x'^2+8x'^2x^2 \right) \sin x.
\eea
Since $A_1(x,-x’)=-A_1(x,x’)$ and $B_1(x,-x’)=B_1(x,x’)$, the equation above can be further written as
\bea \label{E_AB_ex}
\Delta E^{\mathrm{cross}}_{AB}(r)
\nonumber&=&-\frac{4G^{2}}{9\pi^{2}(\omega_{A}+\omega_{B})r^{8}}\int_{0}^{+\infty} d\omega\omega\frac{\hat{\alpha}_{A}\hat{\chi}_{B}(\omega_{A}+\omega_{B}+\omega)}{(\omega_{A}+\omega)(\omega_{B}+\omega)}  \\
&&\times\int_{-\infty}^{+\infty}d{\omega}'\left(\frac{1}{\omega+{\omega}'}-\frac{1}{{\omega}'-\omega}\right){\omega}’\Big[A_{1}(\omega r,{\omega}'r)-iB_{1}(\omega r,{\omega}'r)\Big]e^{i{\omega}'r}.
\eea
Then, performing the Cauchy principal value integral over the variable ${\omega}’$ in Eq. (\ref{E_AB_ex}), one obtains
\bea \label{E_AB_Cauchy}
\nonumber
\Delta E^{\mathrm{cross}}_{AB}(r)&=&-\frac{4G^{2}}{9\pi(\omega_{A}+\omega_{B})r^{8}}\int_{0}^{+\infty}d\omega{\omega}^{2}\frac{\hat{\alpha}_{A}\hat{\chi}_{B}(\omega_{A}+\omega_{B}+\omega)}{(\omega_{A}+\omega)(\omega_{B}+\omega)}   \\
&&\times\Big[A_{2}(\omega r)\cos(2\omega r)+B_{2}(\omega r)\sin(2\omega r)\Big],
\eea
where
\beq \label{A_2}
A_{2}(x)=-90x+42x^3-4x^5,
\eeq
\beq \label{B_2}
B_{2}(x)=45-81x^2+14x^4-x^6.
\eeq
Since $A_2(-x)=-A_2(x)$ and $B_2(-x)=B_2(x)$, the equation above can be  further  transformed into
\bea \label{E_AB_A2B2}
\Delta E^{\mathrm{cross}}_{AB}(r)
\nonumber&=&-\frac{2G^{2}}{9\pi(\omega_{A}+\omega_{B})r^{8}}
\Bigg\{\int_{0}^{+\infty}d\omega{\omega}^2\frac{\hat{\alpha}_{A}\hat{\chi}_{B}(\omega_{A}+\omega_{B}+\omega)}{(\omega_{A}+\omega)(\omega_{B}+\omega)}\Big[A_{2}(\omega r)-iB_{2}(\omega r)\Big] e^{i2\omega r}   \\
&&+\int_{0}^{-\infty}d\omega{\omega}^2\frac{\hat{\alpha}_{A}\hat{\chi}_{B}(\omega_{A}+\omega_{B}-\omega)}{(\omega_{A}-\omega)(\omega_{B}-\omega)}\Big[A_{2}(\omega r)-iB_{2}(\omega r)\Big]e^{i2\omega r}\Bigg\}.
\eea
Letting $\omega=iu$ and performing the integral on the imaginary axis, Eq.~(\ref{E_AB_A2B2}) can further be simplified as
\beq \label{E_AB_simp}
\Delta E^{\mathrm{cross}}_{AB}(r)=\frac{4G^{2}}{9\pi r^{8}}\int_{0}^{+\infty}du u^2\alpha_{A}(iu)\chi_{B}(iu)F(ur)e^{-2ur},
\eeq
where
\bea \label{F_ur}
F(x)=45+90x+81x^2+42x^3+14x^4+4x^5+x^6,
\eea
and
\beq \label{alpha_A}
\alpha_{A}(iu)=\lim\limits_{\epsilon\to0^{+}}\frac{\hat{\alpha}_{A}\omega_{A}}{\omega^{2}_{A}-(iu)^{2}-i\epsilon(iu)},
\eeq
\beq \label{chi_B}
\chi_{B}(iu)=\lim\limits_{\epsilon\to0^{+}}\frac{\hat{\chi}_{B}\omega_{B}}{\omega^{2}_{B}-(iu)^{2}-i\epsilon(iu)},
\eeq
denote the ground-state gravitoelectric and gravitomagnetic polarizabilities of the objects, respectively, which satisfy the following relations: 
\beq \label{alpha_A(B)}
Q_{ij}(iu)=\alpha(iu)E_{ij}(iu,\textbf{r}),
\eeq
\beq \label{chi_A(B)}
S_{ij}(iu)=\chi(iu)B_{ij}(iu,\textbf{r}).
\eeq

Now, we examine the distance dependence of the quantum gravitoelectric-gravitomagnetic interaction energy (\ref{E_AB_simp}) in both the retarded and nonretarded regimes. In the retarded regime, i.e., $r\gg\omega^{-1}_{A(B)}$, the exponential decay term in Eq. (\ref{E_AB_simp}) restricts the $u$ integral to a range where the polarizabilities of the two objects $\alpha_{A}(iu)$ and $\chi_{B}(iu)$ are well approximated by their static values $\alpha_{A}(0)$ and $\chi_{B}(0)$, respectively. After performing the integral over the variable $u$, the interaction energy in the retarded regime is found to be 
\beq \label{E_AB, far_SI}
\Delta E^{\mathrm{cross,R}}_{AB}(r)=\frac{187\hbar c G^{2}}{\pi r^{11}}\alpha_{A}(0)\chi_{B}(0).
\eeq
Note that the result is shown in International System of Units (SI units). In the nonretarded regime, i.e., $r\ll\omega^{-1}_{A(B)}$, all terms in $F(x)$ that contain the variable $ur$ can be neglected, and since $e^{-2ur}\simeq 1$, the interaction energy in the nonretarded regime can be expressed as
\beq \label{E_AB, near_SI}
\Delta E^{\mathrm{cross,NR}}_{AB}(r)=\frac{20\hbar G^{2}}{\pi c^2 r^{8}}\int_{0}^{+\infty}du{u}^2\alpha_{A}(iu)\chi_{B}(iu).
\eeq
Based on the definitions of the gravitoelectric and gravitomagnetic polarizabilities in Eqs. (\ref{alpha_A}) and (\ref{chi_B}), the frequency-dependent polarizabilities $\alpha_A(\omega)$ and $\chi_B(\omega)$ are given by
\beq \label{fre-dep-polar1}
 \alpha _{A}(\omega) =\frac{\alpha _{A}(0)}{1-\left( \frac{\omega}{\omega _{A}} \right) ^2},
\eeq
\beq \label{fre-dep-polar2}
 \chi _{B}(\omega) =\frac{\chi _{B}(0)}{1-\left( \frac{\omega}{\omega _{B}} \right) ^2}.
\eeq 
Taking Eqs. (\ref{fre-dep-polar1}) and (\ref{fre-dep-polar2}) into Eq. (\ref{E_AB, near_SI}), and performing the integration over the variable $u$, the explicit expression of the interaction energy in the nonretarded regime reads
\beq \label{new_near_E}
 \Delta E_{AB}^{\mathrm{cross,NR}}\left( r \right)=\frac{10\hbar G^2}{c^2r^{8}}\frac{\omega^2_A\omega^2_B}{\omega_A+\omega_B}\alpha_A\left( 0 \right) \chi _B\left( 0 \right).
\eeq

The results above show that the quantum gravitoelectric-gravitomagnetic interaction between a gravitoelectrically polarizable object and a gravitomagnetically polarizable object induced by fluctuating gravitoelectric and gravitomagnetic fields in vacuum exhibits an $r^{-8}$ dependence in the near regime and an $r^{-11}$ dependence in the far regime. In the retarded regime, the distance dependence of the quantum gravitational interaction from the gravitoelectric-gravitomagnetic cross term follows the same form as those in the gravitoelectric-gravitoelectric and gravitomagnetic-gravitomagnetic cases. This can be seen by comparing Eq.~(\ref{E_AB, far_SI}) in this paper with Eq.~(1) in Ref.~\cite{Ford2016} and Eq.~(46) in Ref.~\cite{hao2024}. However, in the near regime, the quantum gravitoelectric-gravitomagnetic interaction follows a distinct power law of $r^{-8}$, compared to the $r^{-10}$ dependence of the gravitoelectric-gravitoelectric and gravitomagnetic-gravitomagnetic interactions. Interestingly, this is consistent with the electromagnetic case, in which the Casimir-Polder interaction energy between an electrically polarizable atom and a paramagnetically polarizable one in the near and far regimes also differs by $r^3$~\cite{Salam2010,Buhmann2013}.

Furthermore, it is noteworthy that the results shown in Eqs. (\ref{E_AB, far_SI}) and (\ref{new_near_E}) reveal that the quantum gravitoelectric-gravitomagnetic interaction energy is positive, regardless of whether it is in the retarded or nonretarded regime. This indicates that the vacuum fluctuation induced quantum gravitational interaction between a gravitoelectrically and a gravitomagnetically polarizable object always exhibits a repulsive behavior, which is distinct from the attractive nature in the gravitoelectric-gravitoelectric and gravitomagnetic-gravitomagnetic cases reported in Refs. \cite{Ford2016,hao2024}.

In Table \ref{compar}, we summarize the power laws and the attractive or repulsive nature of the quantum gravitoelectric-gravitoelectric (GE-GE), gravitomagnetic-gravitomagnetic (GM-GM), and gravitoelectric-gravitomagnetic (GE-GM) interaction potentials in the asymptotic regimes. 
In the retarded regime, all three interaction potentials exhibit by a $r^{-11}$ power law.  
In the nonretarded regime, the quantum gravitoelectric-gravitoelectric and gravitomagnetic-gravitomagnetic potentials follow a $r^{-10}$ power law, while the crossed gravitoelectric-gravitomagnetic potentials investigated in this work exhibit a weaker $r^{-8}$ dependence. 
For the attractive or repulsive nature, a general rule emerges: interactions between objects  polarizable in the same gravitoelectric or gravitomagnetic manner are always attractive, while crossed interactions are always repulsive. This implies that, for a pair of objects which are both gravitoelectrically and gravitomagnetically polarizable, the overall quantum gravitational interaction potential is reduced when the gravitoelectric-gravitomagnetic cross interaction is taken into account.

\begin{table*}
\centering
\caption{Power laws and attractive/repulsive nature of quantum gravitational quadrupolar interaction potentials}
\begin{threeparttable}
\begin{tabular}{llll} 
\hline
\hline
\toprule
\cline{2-4}  % line2-4
   &    GE-GE~\tnote{1} & GM-GM & GE-GM (cross term) \\
  \midrule
  \hline
  Nonretarded regime~~     &  $-\frac{315\hbar G^2}{2 r^{10}}\frac{\omega _A\omega _B\alpha _A\left( 0 \right) \alpha _B\left( 0 \right)}{\omega _A+\omega _B}$ & $-\frac{280\hbar G^2}{9 r^{10}}\frac{\omega _A\omega _B\chi _A\left( 0 \right) \chi _B\left( 0 \right)}{\omega _A+\omega _B}$ & $ \frac{10\hbar G^2}{c^2r^{8}}\frac{\omega^2_A\omega^2_B\alpha_A\left( 0 \right) \chi _B\left( 0 \right)}{\omega_A+\omega_B}$  \\
  Retarded regime~~ & $-\frac{3987\hbar c G^{2}\alpha_{A}(0)\alpha_{B}(0)}{4\pi r^{11}}$ & $-\frac{1772\hbar c G^{2}\chi_{A}(0)\chi_{B}(0)}{9\pi r^{11}}$ & $ \frac{187\hbar c G^{2}\alpha_{A}(0)\chi_{B}(0)}{\pi r^{11}}$  \\
  Attractive/repulsive~~ & Attractive & Attractive  & Repulsive  \\
  \bottomrule
\hline
\hline
  \end{tabular}
  \label{compar}
  \begin{tablenotes}
        \footnotesize
        \item[1] The terms  ``GE" and ``GM" denote gravitoelectric and gravitomagnetic, respectively.
  \end{tablenotes}
  \end{threeparttable}
\end{table*}

Now, an intriguing question naturally arises as to whether a regime exists in which the repulsive quantum gravitational quadrupole interaction investigated here dominates over the attractive quantum gravitational quadrupole interactions. The answer is that, for two isotropically polarizable objects with identical gravitoelectric and gravitomagnetic polarizabilities and energy level spacing, the repulsive quantum gravitational interaction cannot exceed the attractive one, irrespective of the gravitoelectric and gravitomagnetic polarizabilities and the interobject distance. The explicit proof is provided in Appendix~\ref{3a}.

\section{Summary }
\label{sec_disc}
%\setcounter{equation}{0}
%%%%%%%%%%%%%%%%%%%%%%%%%%%%%%%%%%%%%%%%%%%%%%%%%%

In this paper, we have studied the quantum gravitational interaction between a gravitoelectrically polarizable object and a gravitomagnetically polarizable object induced by fluctuating gravitoelectric and gravitomagnetic fields in a vacuum, within the framework of linearized quantum gravity. This interaction originates from the coupling between the induced instantaneous mass quadrupole moment and mass-current quadrupole moment in nonpointlike objects. Our result shows that, the quantum gravitoelectric-gravitomagnetic interaction exhibits a distance dependence of $r^{-8}$ in the near regime and $r^{-11}$ in the far regime, where $r$ is the distance between the two objects. Furthermore, the quantum gravitoelectric-gravitomagnetic interaction energy is positive both in the near and far regimes, which indicates that the force is repulsive. Since interactions between objects polarizable in the same gravitoelectric or gravitomagnetic manner are inherently attractive, for  objects which are both gravitoelectrically and gravitomagnetically polarizable, the overall quantum gravitational interaction potential is reduced  when the repulsive quantum gravitoelectric-gravitomagnetic interaction is taken into account. However, for two isotropically polarizable objects with identical gravitoelectric and gravitomagnetic polarizabilities and energy level spacing, the repulsive quantum interaction cannot surpass the attractive interactions, regardless of the gravitoelectric and gravitomagnetic polarizabilities and the interobject distance.

\begin{acknowledgments}

We would like to thank the anonymous referee for the insightful comments and helpful suggestions.
This work was supported in part by the NSFC under Grant No. 12075084, and the innovative research group of Hunan Province under Grant No. 2024JJ1006.

\end{acknowledgments}

%\newpage

\appendix
\section{Intermediate processes of the quantum gravitoelectric-gravitomagnetic interaction}\label{appd1}
The twelve possible intermediate states and their corresponding energy denominators $\mathcal{D}_{(n)} (n=1,2,3,...,12)$ in Eq.~(\ref{E_AB}) are as in Table \ref{24I}.

\begin{table}[H]
  \centering
  \scalebox{0.75}{
\begin{tabular}{cllll}
\hline
Case  &\hspace{1ex} $|\text{I}\rangle$ &\hspace{2ex}$|\text{II}\rangle$ &\hspace{2ex}$|\text{III}\rangle$ &\hspace{2ex} Denominator\\
\hline
(1) &\hspace{1ex} $|e_A\rangle |g_B\rangle |1\rangle $
      & \hspace{2ex}$|g_A\rangle |g_B\rangle |1,{1}'\rangle $
      & \hspace{2ex}$|g_A\rangle |e_B\rangle |{1}'\rangle $
      & \hspace{2ex}$\mathcal{D}_{(1)}=({\omega}'+\omega_{B})({\omega}'+\omega)(\omega+\omega_{A})$  \\
(2) &\hspace{1ex} $|e_A\rangle |g_B\rangle |1\rangle$
      & \hspace{2ex}$|g_A\rangle |g_B\rangle |1,{1}'\rangle$
      & \hspace{2ex}$|g_A\rangle |e_B\rangle |1\rangle$
      & \hspace{2ex}$\mathcal{D}_{(2)}=(\omega+\omega_{B})({\omega}'+\omega)(\omega+\omega_{A})$  \\
(3) &\hspace{1ex} $|e_A\rangle |g_B\rangle |1\rangle$
      & \hspace{2ex}$|e_A\rangle |e_B\rangle |0\rangle $
      & \hspace{2ex}$|g_A\rangle |e_B\rangle |{1}'\rangle $
      & \hspace{2ex}$\mathcal{D}_{(3)}=({\omega}'+\omega_{B})(\omega_{B}+\omega_{A})(\omega+\omega_{A})$  \\
(4) &\hspace{1ex} $|e_A\rangle |g_B\rangle |1\rangle$
      & \hspace{2ex}$|e_A\rangle |e_B\rangle |0\rangle$
      & \hspace{2ex}$|e_A\rangle |g_B\rangle |{1}'\rangle $
      & \hspace{2ex}$\mathcal{D}_{(4)}=({\omega}'+\omega_{A})(\omega_{B}+\omega_{A})(\omega+\omega_{A})$  \\
(5) &\hspace{1ex} $|e_A\rangle |g_B\rangle |1\rangle$
      & \hspace{2ex}$|e_A\rangle |e_B\rangle |1,{1}'\rangle $
      & \hspace{2ex}$|g_A\rangle |e_B\rangle |1\rangle $
      & \hspace{2ex}$\mathcal{D}_{(5)}=(\omega+\omega_{B})(\omega_{B}+\omega_{A}+{\omega}'+\omega)(\omega+\omega_{A})$  \\
(6) &\hspace{1ex} $|e_A\rangle |g_B\rangle |1\rangle$
      & \hspace{2ex}$|e_A\rangle |e_B\rangle |1,{1}'\rangle$
      & \hspace{2ex}$|e_A\rangle |g_B\rangle |{1}'\rangle $
      & \hspace{2ex}$\mathcal{D}_{(6)}=({\omega}'+\omega_{A})(\omega_{B}+\omega_{A}+{\omega}'+\omega)(\omega+\omega_{A})$  \\
(7) &\hspace{1ex} $|g_A\rangle |e_B\rangle |1\rangle $
      & \hspace{2ex}$|g_A\rangle |g_B\rangle |1,{1}'\rangle $
      & \hspace{2ex}$|e_A\rangle |g_B\rangle |{1}'\rangle $
      & \hspace{2ex}$\mathcal{D}_{(7)}=({\omega}'+\omega_{A})({\omega}'+\omega)(\omega+\omega_{B})$  \\
(8) &\hspace{1ex} $|g_A\rangle |e_B\rangle |1\rangle$
      & \hspace{2ex}$|g_A\rangle |g_B\rangle |1,{1}'\rangle$
      & \hspace{2ex}$|e_A\rangle |g_B\rangle |1\rangle$
      & \hspace{2ex}$\mathcal{D}_{(8)}=(\omega+\omega_{A})({\omega}'+\omega)(\omega+\omega_{B})$  \\
(9) &\hspace{1ex} $|g_A\rangle |e_B\rangle |1\rangle$
      & \hspace{2ex}$|e_A\rangle |e_B\rangle |0\rangle $
      & \hspace{2ex}$|e_A\rangle |g_B\rangle |{1}'\rangle $
      & \hspace{2ex}$\mathcal{D}_{(9)}=({\omega}'+\omega_{A})(\omega_{B}+\omega_{A})(\omega+\omega_{B})$  \\
(10)&\hspace{1ex} $|g_A\rangle |e_B\rangle |1\rangle$
      & \hspace{2ex}$|e_A\rangle |e_B\rangle |0\rangle$
      & \hspace{2ex}$|g_A\rangle |e_B\rangle |{1}'\rangle $
      & \hspace{2ex}$\mathcal{D}_{(10)}=({\omega}'+\omega_{B})(\omega_{B}+\omega_{A})(\omega+\omega_{B})$  \\
(11)&\hspace{1ex} $|g_A\rangle |e_B\rangle |1\rangle$
      & \hspace{2ex}$|e_A\rangle |e_B\rangle |1,{1}'\rangle$
      & \hspace{2ex}$|e_A\rangle |g_B\rangle |1\rangle $
      & \hspace{2ex}$\mathcal{D}_{(11)}=(\omega+\omega_{A})(\omega_{B}+\omega_{A}+{\omega}'+\omega)(\omega+\omega_{B})$  \\
(12)&\hspace{1ex} $|g_A\rangle |e_B\rangle |1\rangle$
      & \hspace{2ex}$|e_A\rangle |e_B\rangle |1,{1}'\rangle$
      & \hspace{2ex}$|g_A\rangle |e_B\rangle |{1}'\rangle$
      & \hspace{2ex}$\mathcal{D}_{(12)}=({\omega}'+\omega_{B})(\omega_{B}+\omega_{A}+{\omega}'+\omega)(\omega+\omega_{B})$  \\
\hline
\end{tabular}}
  \caption{Twelve intermediate states of the interaction energy and the expressions of their corresponding energy denominators.}\label{24I}
\end{table}

\section{Summation of gravitoelectric-gravitomagnetic polarization tensors and the symmetry of the two-point functions}\label{appd2}

Let us introduce a triad of coordinate-independent orthogonal unit vectors, denoted as $[\textbf{e}_{1}(\textbf{k}),\textbf{e}_{2}(\textbf{k}),\textbf{e}_{3}(\textbf{k})]$. 
The vector $\textbf{e}_{3}(\textbf{k})=\textbf{k}/k\equiv\hat{k}$ represents the unit vector in the direction of the gravitational field's propagation. The orthogonal relations and the cross product relations satisfied by this triad can be written in the coordinate system which describes the spacetime metric as
\beq \label{orth_ 2}
e^{i}_{a}(\textbf{k})e^{j}_{a}(\textbf{k})=e^{i}_{1}e^{j}_{1}+e^{i}_{2}e^{j}_{2}+\hat{k}^{i}\hat{k}^{j}=\delta_{ij}, 
\eeq
\beq \label{cro_ pro_1}
\epsilon_{ijk}e^{j}_{3}e^{k}_{1}=e^{i}_{2},
\eeq
\beq \label{cro_ pro_2}
\epsilon_{ijk}e^{j}_{3}e^{k}_{2}=-e^{i}_{1},
\eeq
and
\beq \label{pro_3}
e^{i}_{1}e^{j}_{2}-e^{i}_{2}e^{j}_{1}=\epsilon_{ijk}e^{k}_{3}=\epsilon_{ijk}\hat{k}_k.
\eeq
Here, $a=1,2,3$, and $i,j=x,y,z$, the $\hat{k}_{i}$ is the $i$th coordinate component of the $\hat{k}$.

With the help of the vectors $\textbf{e}_{1}(\textbf{k})$ and $\textbf{e}_{2}(\textbf{k})$ in the triad, the gravitational polarization tensor $e_{ij}(\textbf{k},\lambda)$ in the transverse-traceless (TT) gauge can be expressed as~\cite{MTW}
\beq \label{polar tensors_1}
e^{ij}(\textbf{k},+)=e^{i}_{1}(\textbf{k})\otimes e^{j}_{1}(\textbf{k})-e^{i}_{2}(\textbf{k})\otimes e^{j}_{2}(\textbf{k}),
\eeq
\beq \label{polar tensors_2}
e^{ij}(\textbf{k},\times)=e^{i}_{1}(\textbf{k})\otimes e^{j}_{2}(\textbf{k})+e^{i}_{2}(\textbf{k})\otimes e^{j}_{1}(\textbf{k}).
\eeq
Thus, the summation of the polarization tensors gives 
\bea \label{sum_polar tensors}
\nonumber
\sum_{\lambda}e_{ij}(\textbf{\textbf{k}},\lambda)e_{kl}(\textbf{k},\lambda)&=&e^{ij}(\textbf{k},+)e^{kl}({\textbf{k},+})+e^{ij}(\textbf{k},\times)e^{kl}(\textbf{k},\times)  \\
\nonumber&=&\left[e^{i}_{1}(\textbf{k})\otimes e^{j}_{1}(\textbf{k})-e^{i}_{2}(\textbf{k})\otimes e^{j}_{2}(\textbf{k})\right]\left[e^{k}_{1}(\textbf{k})\otimes e^{l}_{1}(\textbf{k})-e^{k}_{2}(\textbf{k})\otimes e^{l}_{2}(\textbf{k})\right]   \\
\nonumber&&+\left[e^{i}_{1}(\textbf{k})\otimes e^{j}_{2}(\textbf{k})+e^{i}_{2}(\textbf{k})\otimes e^{j}_{1}(\textbf{k})\right]\left[e^{k}_{1}(\textbf{k})\otimes e^{l}_{2}(\textbf{k})+e^{k}_{2}(\textbf{k})\otimes e^{l}_{1}(\textbf{k})\right].  \\
\eea

Based on Eq. (\ref{sum_polar tensors}), the gravitoelectric-gravitomagnetic polarization summation term $\mathcal{G}^{\mathrm{E-M}}_{ijab}(\textbf{k})$ in Eq.~(\ref{g_ijab_cal}) can be further expressed as
\bea \label{re_polar_sum}
\nonumber
\mathcal{G}^{\mathrm{E-M}}_{ijab}(\textbf{k})&=&\epsilon_{apq}e^{p}_{3}\sum_{\lambda}e_{ij}(\textbf{\textbf{k}},\lambda)e_{qb}(\textbf{k},\lambda)   \\
\nonumber&=&\epsilon_{apq}e^{p}_{3}\Big\{\left[e^{i}_{1}(\textbf{k})\otimes e^{j}_{1}(\textbf{k})-e^{i}_{2}(\textbf{k})\otimes e^{j}_{2}(\textbf{k})\right]\left[e^{q}_{1}(\textbf{k})\otimes e^{b}_{1}(\textbf{k})-e^{q}_{2}(\textbf{k})\otimes e^{b}_{2}(\textbf{k})\right] \\
\nonumber&&+\left[e^{i}_{1}(\textbf{k})\otimes e^{j}_{2}(\textbf{k})+e^{i}_{2}(\textbf{k})\otimes e^{j}_{1}(\textbf{k})\right]\left[e^{q}_{1}(\textbf{k})\otimes e^{b}_{2}(\textbf{k})+e^{q}_{2}(\textbf{k})\otimes e^{b}_{1}(\textbf{k})\right]\Big\}  \\
\nonumber&=&\left[ e_{1}^{i}\left( \textbf{k} \right) \otimes e_{1}^{j}\left( \textbf{k} \right) -e_{2}^{i}\left( \textbf{k} \right) \otimes e_{2}^{j}\left( \textbf{k} \right) \right]  \\
\nonumber&&\times\left\{ \left[ \epsilon _{apq}e_{3}^{p}e_{1}^{q} \left( \textbf{k} \right) \right] \otimes e_{1}^{b}\left( \textbf{k} \right) -\left[\epsilon _{apq}e_{3}^{p}e_{2}^{q}\left( \textbf{k} \right) \right] \otimes e_{2}^{b}\left( \textbf{k} \right)  \right\}   \\
\nonumber&&+\left[ e_{1}^{i}\left( \textbf{k} \right) \otimes e_{2}^{j}\left( \textbf{k} \right) +e_{2}^{i}\left( \textbf{k} \right) \otimes e_{1}^{j}\left( \textbf{k} \right) \right]   \\
&&\times\left\{ \left[ \epsilon _{apq}e_{3}^{p}e_{1}^{q}\left( \textbf{k} \right)\right] \otimes e_{2}^{b}\left( \textbf{k} \right) +\left[\epsilon _{apq}e_{3}^{p}e_{2}^{q}\left( \textbf{k} \right)\right] \otimes e_{1}^{b}\left( \textbf{k} \right) \right\}.
\eea
Then, using the cross product relations shown in Eqs. (\ref{cro_ pro_1}) and (\ref{cro_ pro_2}), the equation above can be simplified as
\bea \label{sum_polar_final}
\nonumber
\mathcal{G}^{\mathrm{E-M}}_{ijab}(\textbf{k})&=&\left[e^{i}_{1}\otimes e^{j}_{1}-e^{i}_{2}\otimes e^{j}_{2} \right]\left[e^{a}_{2}\otimes e^{b}_{1}+e^{a}_{1}\otimes e^{b}_{2} \right]   \\
&&+\left[e^{i}_{1}\otimes e^{j}_{2}+e^{i}_{2}\otimes e^{j}_{1} \right]\left[e^{a}_{2}\otimes e^{b}_{2}-e^{a}_{1}\otimes e^{b}_{1} \right].
\eea

Furthermore, by utilizing Eqs. (\ref{orth_ 2}) and (\ref{pro_3}), the equation above can further be calculated as follows
\bea \label{sum_result}
\nonumber
\mathcal{G}^{\mathrm{E-M}}_{ijab}(\textbf{k})&=&e^{i}_{1}e^{j}_{1}e^{a}_{2}e^{b}_{1}+e^{i}_{1}e^{j}_{1}e^{a}_{1}e^{b}_{2}-e^{i}_{2}e^{j}_{2}e^{a}_{2}e^{b}_{1}-e^{i}_{2}e^{j}_{2}e^{a}_{1}e^{b}_{2}   \\
\nonumber&&+e^{i}_{1}e^{j}_{2}e^{a}_{2}e^{b}_{2}-e^{i}_{1}e^{j}_{2}e^{a}_{1}e^{b}_{1}+e^{i}_{2}e^{j}_{1}e^{a}_{2}e^{b}_{2}-e^{i}_{2}e^{j}_{1}e^{a}_{1}e^{b}_{1}   \\
\nonumber&=&\left( e^{i}_{1}e^{a}_{2}-e^{i}_{2}e^{a}_{1}  \right)\left( e^{j}_{1}e^{b}_{1}+e^{j}_{2}e^{b}_{2} \right)+\left( e^{i}_{1}e^{a}_{1}+e^{i}_{2}e^{a}_{2} \right)\left( e^{j}_{1}e^{b}_{2}-e^{j}_{2}e^{b}_{1} \right)  \\
\nonumber&=&\epsilon_{iam}\hat{k}_{m}\left( \delta_{jb}-\hat{k}_{j}\hat{k}_{b} \right)+\left( \delta_{ia}-\hat{k}_{i}\hat{k}_{a} \right)\epsilon_{jbn}\hat{k}_{n} \\
&=&\epsilon_{iam}\hat{k}_m\delta_{jb}-\epsilon_{iam}\hat{k}_m\hat{k}_j\hat{k}_b+\delta_{ia}\epsilon _{jbn}\hat{k}_n-\hat{k}_i\hat{k}_a\epsilon _{jbn}\hat{k}_n.
\eea

Based on the analysis above, let us consider the form of the two-point functions when the order of the gravitoelectric and the gravitomagnetic field operators is exchanged. In this case, it becomes
\bea \label{sy_pr}
\nonumber
G^{\mathrm{M-E}}_{ijab}(\textbf{r},{\textbf{r}}',t,{t}')&=&\langle 0|B_{ij}(\textbf{r},t)E_{ab}(\textbf{r}',{t}')|0\rangle  \\
\nonumber&=&\sum_{\lambda}\int d^{3}\textbf{k}\int d^{3}{\textbf{k}}'\left(\frac{G\sqrt{\omega^{3}{\omega}'^{3}}}{(2\pi)^{2}}\right)\epsilon_{ifl}e^{f}_{3}e_{lj}(\textbf{k},\lambda)e_{ab}({\textbf{k}}',\lambda)  \\
\nonumber&&\times\biggl\langle 0\bigg|\left[ g_{\textbf{k},\lambda}e^{i\textbf{k}\cdot\textbf{r}-i\omega t}+g^{\dagger}_{\textbf{k},\lambda}e^{-i\textbf{k}\cdot\textbf{r}+i\omega t}\right]  \\
\nonumber&&\times\left[ g_{\textbf{k}',\lambda}e^{i{\textbf{k}}'\cdot{\textbf{r}}'-i{\omega}' {t}'}+g^{\dagger}_{\textbf{k}',\lambda}e^{-i{\textbf{k}}'\cdot{\textbf{r}}'+i{\omega}' {t}'}\right]\bigg|0\biggr\rangle  \\
&=&\int d^{3}\textbf{k}\frac{G\omega^{3}}{(2\pi)^{2}}\mathcal{G}^{\mathrm{M-E}}_{ijab}(\textbf{k})e^{i\textbf{k}\cdot(\textbf{r}-{\textbf{r}}')-i\omega(t-{t}')},
\eea
where, $\mathcal{G}^{\mathrm{M-E}}_{ijab}(\textbf{k})=\sum_{\lambda}\epsilon_{ifl}e^{f}_{3}e_{lj}(\textbf{k},\lambda)e_{ab}(\textbf{k},\lambda)$ is the gravitomagnetic-gravitoelectric polarization summation term, whose explicit expression is
\bea \label{sum_result_ex}
\nonumber
\mathcal{G}^{\mathrm{M-E}}_{ijab}(\textbf{k})&=&-e^{i}_{1}e^{j}_{1}e^{a}_{2}e^{b}_{1}-e^{i}_{1}e^{j}_{1}e^{a}_{1}e^{b}_{2}+e^{i}_{2}e^{j}_{2}e^{a}_{2}e^{b}_{1}+e^{i}_{2}e^{j}_{2}e^{a}_{1}e^{b}_{2}   \\
\nonumber&&-e^{i}_{1}e^{j}_{2}e^{a}_{2}e^{b}_{2}+e^{i}_{1}e^{j}_{2}e^{a}_{1}e^{b}_{1}-e^{i}_{2}e^{j}_{1}e^{a}_{2}e^{b}_{2}+e^{i}_{2}e^{j}_{1}e^{a}_{1}e^{b}_{1}   \\
\nonumber&=&-\epsilon_{iam}\hat{k}_m\delta_{jb}+\epsilon_{iam}\hat{k}_m\hat{k}_j\hat{k}_b-\delta_{ia}\epsilon _{jbn}\hat{k}_n+\hat{k}_i\hat{k}_a\epsilon _{jbn}\hat{k}_n  \\
&=&-\mathcal{G}^{\mathrm{E-M}}_{ijab}(\textbf{k}),
\eea
which can be derived using the methods shown in Eqs. (\ref{re_polar_sum})-(\ref{sum_result}). 
Therefore, we have
\bea
    \label{symmetry_proof}
    G^{\mathrm{E-M}}_{ijkl}(\textbf{r},{\textbf{r}}',t,t')=-G^{\mathrm{M-E}}_{ijkl}(\textbf{r},{\textbf{r}}',t,t').
\eea

\section{Comparison of the magnitudes of the repulsive quantum gravitoelectric-gravitomagnetic interaction and the attractive gravitoelectric-gravitoelectric and gravitomagnetic-gravitomagnetic interactions }\label{3a}

To compare the magnitude of the repulsive quantum gravitoelectric-gravitomagnetic interaction with those of the attractive gravitoelectric-gravitoelectric and gravitomagnetic-gravitomagnetic interactions, we start from the original integral expressions for the interaction potentials, as presented in Eq. (\ref{E_AB_simp}) of this paper, Eq. (28) in Ref. \cite{Ford2016}, and Eq. (39) in Ref. \cite{hao2024}. For two isotropically polarizable objects which are both gravitoelectrically and gravitomagnetically polarizable, we assume, for convenience, that their gravitoelectric and gravitomagnetic polarizabilities are equal, and their  energy level spacings are the same, i.e., $\hat{\alpha}_A=\hat{\alpha}_B=\alpha$, $\hat{\chi}_A=\hat{\chi}_B=\chi$, and  $\omega_A=\omega_B=\omega_0$. Then, with the help of Eqs. (\ref{alpha_A}) and (\ref{chi_B}), the original integral expressions for the three types of potentials can be reorganized as
\beq \label{orig_em}
\Delta E_{AB}^{\mathrm{cross}}\left( r \right) =\frac{G^2\omega _{0}^{2}r^4}{\pi r^{11}}\int_0^{+\infty}{dx\,\frac{4}{9}\alpha \chi T_1 \left( x \right) \left[ \frac{1}{\left( \omega _0r \right) ^2+x^2} \right] ^2e^{-2x}},
\eeq
\beq \label{orig_ee}
\Delta E_{AB}^{\mathrm{GE}-\mathrm{GE}}\left( r \right) =-\frac{G^2\omega _{0}^{2}r^4}{\pi r^{11}}\int_0^{+\infty}{dx\,\alpha ^2T_2\left( x \right) \left[ \frac{1}{\left( \omega _0r \right) ^2+x^2} \right] ^2e^{-2x}},
\eeq
and
\beq \label{orig_mm}
\Delta E_{AB}^{\mathrm{GM}-\mathrm{GM}}\left( r \right) =-\frac{G^2\omega _{0}^{2}r^4}{\pi r^{11}}\int_0^{+\infty}{dx\,\frac{16}{81}\chi ^2T_2\left( x \right) \left[ \frac{1}{\left( \omega _0r \right) ^2+x^2} \right] ^2e^{-2x}},
\eeq
respectively. Note that the results above are shown in natural units, with $x=ur$, and $T_1(x)$ and $T_2(x)$ take the forms
\beq \label{T1_x}
T_1\left( x \right) =x^2\left( 45+90x+81x^2+42x^3+14x^4+4x^5+x^6 \right),
\eeq
and
\beq \label{T2_x}
T_2\left( x \right) =315+630x+585x^2+330x^3+129x^4+42x^5+14x^6+4x^7+x^8,
\eeq
respectively. Then, the magnitudes of the attractive and repulsive quantum gravitational interactions are given by $|\Delta E_{AB}^{\mathrm{GE}-\mathrm{GE}}\left( r \right)|+|\Delta E_{AB}^{\mathrm{GM}-\mathrm{GM}}\left( r \right)|$ and $|\Delta E_{AB}^{\mathrm{cross}}\left( r \right)|$, respectively, where ``$|\,\,|$’’ denotes the absolute value. In the following, we prove that their difference,
\beq \label{DD}
\Delta E =|\Delta E_{AB}^{\mathrm{GE}-\mathrm{GE}}\left( r \right)|+|\Delta E_{AB}^{\mathrm{GM}-\mathrm{GM}}\left( r \right)|-|\Delta E_{AB}^{\mathrm{cross}}\left( r \right)|,
\eeq
is always positive, indicating that the repulsive quantum gravitoelectric-gravitomagnetic interaction cannot dominate.

Substituting Eqs. (\ref{orig_em})-(\ref{orig_mm}) into Eq. (\ref{DD}), then it can be shown that $\Delta E > 0$ implies
\beq \label{bijiao}
\int_0^{+\infty}{dx\left\{ \left[ \alpha ^2T_2\left( x \right) +\frac{16}{81}\chi ^2T_2\left( x \right) \right] -\left[ \frac{4}{9}\alpha \chi T_1\left( x \right) \right] \right\} \left[ \frac{1}{\left( \omega _0r \right) ^2+x^2} \right] ^2e^{-2x}}>0.
\eeq
According to Eqs. (\ref{polar_1}), (\ref{polar_2}), and (\ref{T2_x}), since $\alpha$, $\chi$, and $T_2(x)$ are all positive, it follows from the inequality of arithmetic and geometric means that
\beq \label{basic_ineq}
\alpha ^2T_2\left( x \right) +\frac{16}{81}\chi ^2T_2\left( x \right) \geq \frac{8}{9}\alpha \chi T_2\left( x \right).
\eeq
Therefore, a sufficient condition for inequality (\ref{bijiao}) to hold is
\beq \label{SSSS}
\delta\equiv\int_0^{+\infty}{dx\left[ \frac{8}{9}T_2\left( x \right) -\frac{4}{9}T_1\left( x \right) \right] \left[ \frac{1}{\left( \omega _0r \right) ^2+x^2} \right] ^2e^{-2x}}>0,
\eeq
which means that the condition $\Delta E > 0$ describing the magnitude of the repulsive interaction energy as being smaller than the attractive interaction energy translates into $\delta > 0$.

Now, let us discuss whether $\delta>0$ is valid in both the nonretarded and retarded regimes, as well as in the transition regime where the interobject distance is comparable to the transition wavelength.

\begin{figure}[ht]
\includegraphics[scale=0.6]{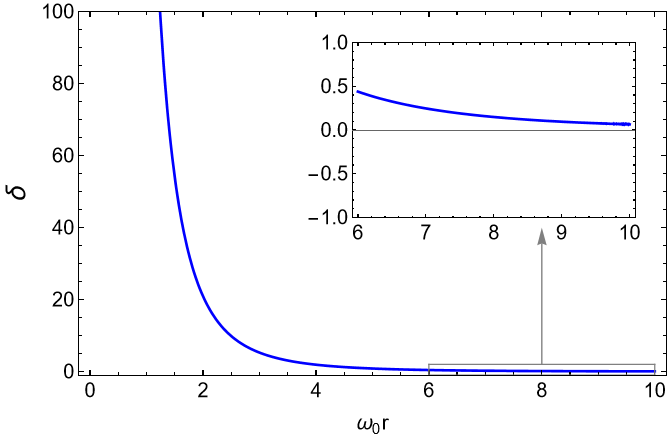}\vspace{0.0cm}
\caption{\label{Delta_1} 
The value of $\delta$ as a function of $\omega_0r$ in the transition regimes between the nonretarded and retarded regimes.}
\end{figure}

(1) In the nonretarded regime, 
performing the integral in Eq. (\ref{SSSS}) and then expanding the result in a Taylor series in the region where $\omega_0 r \ll 1$, we obtain
\beq \label{near_com}
\delta =\frac{70\pi}{\left( \omega _0r \right) ^3}+\mathcal{O} \left[ \frac{1}{\left( \omega _0r \right)} \right] ,
\eeq
which is positive. Here, $\mathcal{O}(x^n)$ denotes terms on the order of $x^n$ or higher, which are considered negligibly small and thus omitted.

(2) In the retarded regime, 
we perform the integral in Eq. (\ref{SSSS}) and then expand the result in a Taylor series in the region where $\omega_0 r \gg 1$, yielding
\beq \label{far_com}
\delta =\frac{699}{\left( \omega _0r \right) ^4}+\mathcal{O} \left[ \frac{1}{\left( \omega _0r \right)^6} \right],
\eeq
which is also positive.

(3) In the transition regime between the nonretarded and retarded regimes, we show numerically the value of $\delta$ as a function of $\omega_0r$ in these transition regimes in Fig.~{\ref{Delta_1}}, which clearly shows that $\delta>0$ always holds.

Considering the analysis above, for two isotropically polarizable objects which are both gravitoelectrically and gravitomagnetically polarizable and have the same gravitoelectric and gravitomagnetic polarizabilities and energy level spacing, the repulsive interaction energy cannot dominate over the attractive one, irrespective of the gravitoelectric and gravitomagnetic polarizabilities and the interobject distance.

\end{document}